%% file: entity_attr.tex
\documentclass[sigconf]{acmart}

\usepackage{booktabs} 
\usepackage{subfigure}
\usepackage{algpseudocode}
\usepackage{multirow}
\usepackage{amsmath}
\usepackage[normalem]{ulem}
\usepackage{wasysym}
\usepackage{balance}
\usepackage{enumitem}
\usepackage{algorithm2e}

\newcommand{\miniskip}{\vspace*{-.5\baselineskip}}
\newcommand{\shrink}{\vspace*{-0.7\baselineskip}}

\newtheoremstyle{mydef}
{2ex}
{2ex}
{\itshape}
{}
{\scshape}
{: }
{0.5em}
{}
\theoremstyle{mydef}
\newtheorem{mydef}{Definition}

\newtheoremstyle{remark}
{1ex}
{1ex}
{\normalfont}
{}
{\scshape}
{: }
{0.5em}
{}
\theoremstyle{remark}
\newtheorem{remark}{Remark}

\copyrightyear{2018}
\acmYear{2018}
\setcopyright{acmcopyright}
\acmConference[CIKM '18]{The 27th ACM International Conference on Information and Knowledge Management}{October 22--26, 2018}{Torino, Italy}
\acmBooktitle{The 27th ACM International Conference on Information and Knowledge Management (CIKM '18), October 22--26, 2018, Torino, Italy}
\acmPrice{15.00}
\acmDOI{10.1145/3269206.3269245}
\acmISBN{978-1-4503-6014-2/18/10}

\acmPrice{15.00}

\begin{document}
\title{Query Understanding via Entity Attribute Identification}

\author{Arash Dargahi Nobari}
\affiliation{%
	\institution{Faculty of Computer Science and Engineering \\Shahid Beheshti University, G.C.}
}
\email{a.dargahinobari@mail.sbu.ac.ir}

\author{Arian Askari}
\affiliation{%
	\institution{Faculty of Computer Science and Engineering \\Shahid Beheshti University, G.C.}
}
\email{ar.askari@mail.sbu.ac.ir}

\author{Faegheh Hasibi}
\affiliation{%
	\institution{Norwegian University of Science and Technology}
}
\email{faegheh.hasibi@ntnu.no}

\author{Mahmood Neshati}
\affiliation{%
	\institution{Faculty of Computer Science and Engineering \\Shahid Beheshti University, G.C.}
}
\email{m_neshati@sbu.ac.ir}


\begin{abstract}
Understanding searchers' queries is an essential component of semantic search systems. 
 In many cases, search queries involve specific attributes of an entity in a knowledge base (KB), which can be further used to find query answers. In this study, we aim to move forward the understanding of queries by identifying their related entity attributes from a knowledge base.  To this end, we introduce the task of entity attribute identification and propose two methods to address it: (i) a model based on Markov Random Field, and (ii) a learning to rank model.
  We develop a human annotated test collection and show that our proposed methods can bring significant improvements over the baseline methods.
 
%
%
%
%
\end{abstract}

%
%


\maketitle

\input{entity_attr_01} 
\input{entity_attr_03} 
\input{entity_attr_04} 
\input{entity_attr_05} 
\input{entity_attr_06} 

\miniskip
\bibliographystyle{ACM-Reference-Format}
\bibliography{bibliography}

\end{document}

%% file: entity_attr_01.tex
\section{Introduction}
\label{sec:intro}

Understanding the underlying intent of search queries is a crucial component in virtually every semantic search system, either being a web search engine, a chatbot, or an e-commerce website. 
It has been long recognized that knowledge bases such as DBpedia,  Freebase, and YAGO are rich sources of information for interpreting and understanding queries.
A large body of efforts in this area is focused on recognizing mentioned entities in the queries and linking them to the corresponding entities in a knowledge base, the so-called task of entity linking in queries~\citep{hasibi:2015:ELQ, Blanco:2015:FSE}.
In this paper, we aim to further the understanding of queries by identifying their entity attributes from a knowledge base; e.g., identifying the attribute \emph{spouse} from DBpedia for the query ``the wife of Lincoln.''

Extracting entity attributes of queries is beneficial for answering the queries in tasks such as question answering and entity retrieval. 
It has been shown that joint entity linking and attribute identification of queries improves question answering over knowledge bases~\citep{Xu:2016:QAF, Yao:2014:IES}. 
Similarly, entity retrieval approaches can benefit from entity attribute identification by having a  focused selection of entity attributes~\citep{Hasibi:2016:ELR} and using them to build fielded representation of entities~\citep{Hasibi:2017:DTC}. 
Entity attribute identification can be also employed in the e-commerce websites to improve search results and boost sites' advertising profits and recommendation quality. Consider, for example, the query ``nike shoes size 38'', where the attribute \emph{size} can be used to filter out irrelevant products or advertising similar products from other brands.

Motivated by the above reasons, we set out to focus on identifying entity attributes that help answering a query. We note that this is a highly non-trivial task, mainly due to vocabulary mismatch between query terms and the entity attribute(s) pointed by the query. Take for example the query ``the father of integrated circuits'', which refers to the attribute \emph{inventor}, rather than \emph{father} or \emph{parent}.
We frame the entity identification task as a ranking problem and propose a set of methods to address it. Our first model is based on Markov Random Field (MRF) and incorporates entity annotations of queries as a bridge to rank entity attributes. We further employ a learning to rank approach combining various attribute similarity scores and show significant improvements with respect to our best baseline. We evaluate our results on a purpose-built test collection based on the DBpedia-Entity v2 collection~\citep{Hasibi:2017:DTC} for entity retrieval.

To summarize, the contributions of this work are as follows:

\begin{itemize}
	\item  We introduce and formulate the task of ``entity attribute identification.'' 
	\item We propose a set of methods (an MRF-based and a learning to rank model) to address the entity attribute identification task, and provide insights into the influence of different contributors of our models.
	\item In order to evaluate the task and foster research in this area, we build a test collection, consisting of graded scores for a diverse set of entity oriented queries. The dataset is human-annotated and is made publicly available at \textcolor{blue}{\url{http://tiny.cc/eai}}.

	\end{itemize}

%% file: entity_attr_03.tex
\miniskip
\section{Entity Attribute Identification}
\label{sec:en_att}
In this section, we formally define the problem of entity attribute identification and describe our proposed methods.
%
\input{entity_attr_03_01}

\input{entity_attr_03_02}
\input{entity_attr_03_03}

%% file: entity_attr_03_01.tex
\miniskip
\subsection{Problem Definition}
\label{sec:en_att:def}

\begin{mydef}[\textbf{Entity attribute identification}]
	Given an entity-bearing query, entity attribute identification is defined as the task of returning a ranked list of entity attributes, where the values of those attributes provide  answers to the query or help finding the answers.
\end{mydef}

\begin{remark}
	In this definition, we focus on entity-bearing queries; i.e., queries that refer to specific entities in a knowledge base. For example the query ``the wife of Lincoln,'' which can be linked to entity \textsc{Abraham Lincoln}. Entity linking in queries is a well studied task, and can be performed using publicly available entity linkers such as TAGME~\cite{tagme} and Nordlys~\cite{Hasibi:2017:NTE}.
\end{remark}

\begin{remark}
	Each entity in knowledge base is represented by a list of pairs \texttt{$e = \{\langle a_1,v_1 \rangle, \langle a_2,v_2 \rangle, ..., \langle a_n,v_n \rangle \}$}, where $a_i$ is an attribute and $v_i$ is its associated value. For example,  the entity \textsc{Abraham Lincoln} is represented as
	\texttt{\{$\langle$spouse, Mary Todd Lincoln$\rangle$, $\langle$death Place, Washington D.C.$\rangle$, ... \}}.
\end{remark}

\begin{remark}
	The ranked entity attributes belong to the entities that are linked to the query.  For example \texttt{\{spouse\}} is the top-ranked attribute for the aforementioned query.
\end{remark}

Here, we relate this problem to the extensive body of work on attribute extraction on (semi-)structured text~\citep{Hoffmann:2010:LRE, Bing:2013:WEE, Zhong:2016:EAN} and highlight that our goal is to further machine-understanding of \emph{queries}, which are short, ambiguous pieces of text (unlike long documents). 
Similar efforts have been performed in e-commerce to extract \emph{attribute values} of product titles~\citep{More:2016:AEP}, and further filter out search results based on the matching attribute values. The most similar task to ours is the NTCIR actionable knowledge graph generation (AKGG) task~\citep{Blanco:2017:ONA}, which aims at ranking attributes of a query that are relevant for performing users' actions. In our task, we consider a different (rather broader) context and identify entity attributes that are useful for finding relevant answers to the query. Consequently, the outcome can be incorporated in various other tasks such as entity retrieval and questions answering.

%% file: entity_attr_03_02.tex
\subsection{MRF-based Model}
\label{sec:prob_model}

Our first model to address entity attribute identification task is based on Markov Random Field (MRF). 
Here, our goal is to compute the relevance probability of an attribute $a$ to a given query $q$, which can be estimated by a set of joint probabilities between the attribute, query, and a linked entity to the query:
%
\begin{equation}
p(a|q) = \frac{p(a,q)}{p(q)} \overset{rank}{=} \sum_{e \in E} p(a,e,q).
\label{eq:first}
\end{equation}
In this equation, $E$ is the set of entities linked to  the query $q$ and is obtained by an entity linker system.

In order to estimate $p(a,e,q)$, we follow the idea of~\citet{Metzler:2005:MRF} in using MRF for ad hoc retrieval tasks. MRF is a graphical model, which can be used for estimating joint probability of random variables described by an undirected graph $G$. In this graph, nodes indicate random variables and edges represent dependency between the nodes. The joint probability over variables of the graph $G$ is computed as: 
%
\begin{equation}
\label{eq:mrf_start}
P_{\Lambda}(G) = \frac{1}{Z_\Lambda} \prod_{c \in C(G)} \psi (c; \Lambda),
\end{equation}
where $C(G)$ is the set of cliques in graph $G$ and $\psi (c; \Lambda) = exp[ \lambda_c f(c)]$ is a non-negative potential function, parametrized by the weight $\lambda_c$ and the feature function $f_c$. The parameter $Z_\Lambda$ is a normalization factor, which is generally ignored due to computational infeasibility. Ignoring  $Z_\Lambda$ and taking logarithm of the right hand side of Eq. \ref{eq:mrf_start}, the joint probability of $P_{\Lambda}(G)$ is proportional to:
\begin{equation}
\label{eq:mrf2}
P_{\Lambda}(G) \propto \sum_{c \in C(G) } \log [\psi (c; \Lambda)] = \sum_{c \in C(G)} \lambda_c f(c).
\end{equation}

\begin{figure}[t]
	\shrink
	\includegraphics[width=.55\linewidth]{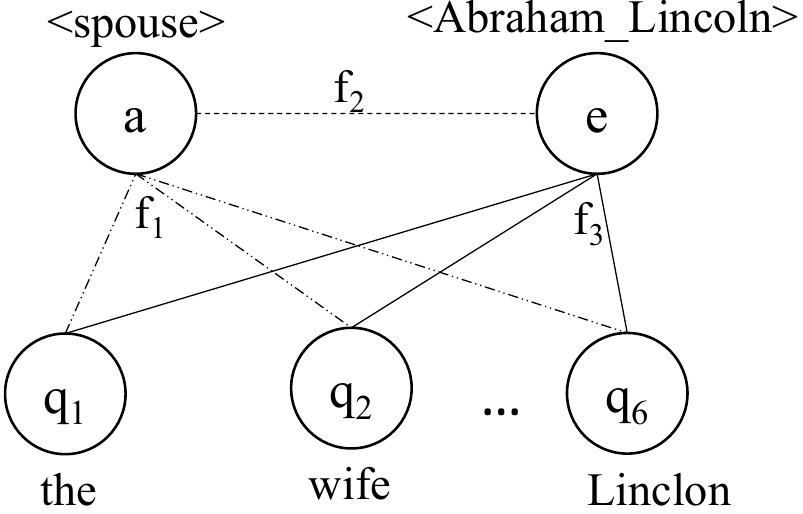}
	\shrink
	\caption{MRF graph for the query ``the wife of U.S. president Lincoln.''}
	\label{fig:graph}
\end{figure}

The graph underlying our model consists of independent query terms, an entity, and an attribute; see Figure~\ref{fig:graph}. 
In this graph, three types of 2-cliques are defined: (i) cliques involving a query term and the attribute, (ii) a clique involving the entity and the attribute, and (iii) cliques involving a query term and the entity. The 3-cliques involving a query term, an attribute, and an entity are ignored due to computational complexity. Putting all these elements together, the probability  $P(a|q)$ is proportional to:
\begin{equation}
\label{eq:ranking}
p(a|q) \overset{rank}{=} \sum_{e \in E} \left(
\lambda_1 \sum_{q_i \in q}{f_1 (q_i,a)} + 
\lambda_2 f_2 (a,e) + 
\lambda_3 \sum_{q_i \in q}{f_3 (q_i,e)}
\right),
\end{equation}
where the $\lambda$ parameters  should meet the constraint of  $\sum_{i=1}^{3} \lambda_i = 1$. 

We now define the feature functions of our model. The  first feature functions is defined as:
\begin{equation}
\label{eq:ff1}
f_1(q_i,a) = log [\frac{1}{|a|} \sum_{w \in a} 1-distance(\vec{q_i},\vec{w})],
\end{equation}
where $w$ is an attribute word and $distance (\vec{q_i},\vec{w})$ indicates the Euclidean distance between the vector representation of words $q_i$ and $w$. We obtain these vector representations form  Word2Vec~\cite{word2vec} 300-dimensions vectors, trained on the Google news dataset. Using this feature function, our model is able to capture the semantic similarity between query and attribute terms; e.g. ``spouse'' and ``wife'' in Fig. \ref{fig:graph}. 

The second feature function is computed by:
\begin{equation}
\label{eq:ff2}
f_2(a,e) = log [
\mu_1 \frac{|\{\langle t,v \rangle  \in e | t=a \}|}{|\{\langle t,v \rangle \in e \}|} + (1-\mu_1) \frac{|\{\langle t,v \rangle \in \mathcal{E} | t=a \}|}{|\{\langle t,v \rangle \in \mathcal{E}\}|}
],
\end{equation}
where $\mu_1$ is the smoothing parameter. Here, $e$ is an entity represented by a set of attribute-value pairs $\langle t,v \rangle$, and  $\mathcal{E}$ is the collection of all these pairs  from all entities in the knowledge base.

The feature function $f_3(q_i,e)$ measures the similarity between an entity and a query term and is defined as:
\begin{align}
\label{eq:ff3}	
f_3(q_i,e) = log [
\mu_2 \frac{|\{\langle t,v \rangle \in e | q_i \in terms(t) \vee q_i \in terms(v) \}|}{|\{\langle t,v \rangle \in e \}|} + \nonumber\\ 
(1-\mu_2) \frac{|\{\langle t,v \rangle \in \mathcal{E} | q_i \in terms(t) \vee q_i \in terms(v)  \}|}{|\{\langle t,v \rangle \in \mathcal{E}\}|}
],  
\end{align}
where $terms(.)$ returns a set of terms of a given text, and $\mu_2$ is a smoothing parameter.


%% file: entity_attr_03_03.tex
\subsection{Learning to Rank Model}
\label{sec:ltr_model}
In this section, we propose a Learning to Rank (LTR) approach for addressing the entity attribute identification task. 
We employ seven features, described in table \ref{tab:features}, and train our learning to rank algorithm. Given the low-dimensional feature space and limited number of training instances, we use Coordinate Ascent (CA)~\citep{metzlerIR07} algorithm for our LTR approach.  



The employed features are as follows. Features $f_1$, $f_2$, and $f_3$ capture entity linking probability, entity attribute similarity, and query attribute similarity (cf. Section \ref{sec:prob_model}). For features $f_4$--$f_7$, we partitioned the query terms into two disjoint sets. The first subset  includes query terms which are linked to an entity (i.e., linked terms) and the second subset  is the set of terms which are not linked to any entity (i.e., not-linked-terms). For example, in the query ``the wife of Lincoln'' linked to entity \textsc{Abraham Lincoln}, the set of linked and not linked terms are \{``Lincoln''\} and \{``the,''``wife,'' ``of''\}, respectively. We then compute the similarity between these terms and concatenation of an attribute-value pair, based on WordNet and Word2Vec~\cite{word2vec} vector representation of words.

\setlength{\abovetopsep}{0pt}
\setlength{\aboverulesep}{0pt}
\setlength{\belowrulesep}{0pt}
\setlength{\belowbottomsep}{0pt}

\begin{table}[t]
  \centering
  \caption{List of features used in LTR approach.}
  \shrink
    \begin{tabular}{|@{~}l@{~}|l@{~}|}
    \hline
    Feature & Description \\
    \midrule
    $f_1$    & $\sum_{q_i \in q}{f_3 (q_i,e)}$  \\
	$f_2$    & $f_2(a,e)$ \\
	$f_3$    & $\sum_{q_i \in q}{f_1 (q_i,a)}$ \\
	$f_4$    & WordNet similarity using linked terms of query $q$  \\
	$f_5$    & Word2Vec similarity using linked terms of query $q$\\
	$f_6$    & WordNet similarity using not linked terms of query $q$ \\
	$f_7$    & Word2Vec similarity using not linked terms of query $q$\\
    \hline
    \end{tabular}%
  \label{tab:features}%
  \vspace*{-1\baselineskip}
\end{table}%

%% file: entity_attr_04.tex
\section{Test Collection Creation}
\label{sec:coll}

In order to evaluate our proposed methods, we created a test collection for the entity attribute identification problem. 
 We used DBpedia 2015-10 as our knowledge base and built our test collection based on DBpedia-Entity v2 collection~\cite{Hasibi:2017:DTC}. 
 This dataset consists of 467 queries and their relevant entities from DBpedia 2015-10. Using DBpedia-Entity v2 collection, we generated a new test collection for the attribute identification task.
 


Our test collection was generated in two steps. 
%
In the first step, we identified all entities that could be linked to the query. To improve recall, we used the two publicly available entity linker systems: TAGME~\cite{tagme} and Nordlys~\cite{Hasibi:2017:NTE}.
 For each entity $e$ linked to query $q$, all its attributes are obtained and added to the pool of candidate attributes if the value of the attribute is among relevant entities of the query $q$.
In the second step, three information retrieval students were asked to annotate query-attribute pairs. They were all trained about the concepts of entities and asked to grade the entity attributes based on the following definitions. These definitions are intentionally inline with the ones from the DBpedia-Entity collection to keep the consistency of datasets.
%
\begin{itemize}
	\item  	\textbf{Highly relevant (2): } The attribute holds direct answer to the user's  query. That is, the attribute should be put among the top results.
	\item  	\textbf{Relevant (1): } The attribute can guide user to find the exact answer, but does not hold direct answer to the query. In other words, the attribute should not be placed among \emph{top} results.
	\item  	\textbf{Irrelevant (0): } The attribute has no relation to the query and should not be considered as an answer.
\end{itemize}

The collection was annotated by three experts, and in case of disagreement the forth annotator was involved. We measured quality of the obtained labels by computing the inter-annotator agreement using Fleiss' Kappa.
Over all candidates, we got an average Kappa of 0.38, which is considered a fair agreement.
The final test collection includes 167 queries and their relevant attributes.
Query categories of our test collection are similar to ones from DBpedia-Entity collection. Table~\ref{tab:collection} summarizes the statistics of the collection.

\begin{table}[t]
  \centering
  \small
  \shrink
 \caption{Query categories in our test collection, QLen indicates the average number of terms per query. $R_1$ and $R_2$ refer to the average number of relevant and highly relevant attributes per query, respectively.}
 \shrink
    \begin{tabular}{|l|l|l|l|l|l|}
    \toprule
    Category & \#queries & QLen  & Type  & $R_1$    & $R_2$ \\
    \midrule
    INEX-LD & 31    & 4.74  & Keyword queries & 2.48  & 1.89 \\
    QALD2 & 62    & 7.52  & NL questions & 2.03  & 2.31 \\
    SemSearch\_ES & 40    & 2.53  & Named entities & 2.94  & 2.18 \\
    ListSerach & 34    & 5.38  & List of entities & 2.38  & 2.38 \\
    \midrule
    Total & 167   & 5.04  &       & 2.46  & 2.19 \\
    \bottomrule
    \end{tabular}%
  \label{tab:collection}%
  \vspace*{-1.4\baselineskip}
\end{table}%

%% file: entity_attr_05.tex

\section{Baselines and  Settings}
For our baseline methods, we ran BM25, Language Model (LM), and Mixture of Language Models (MLM)~\cite{mlm} on an index built based on DBpedia. 
Each document in our index is identified by an entity-attribute pair. Considering the entity $e$ with $k$ attributes $\{a_1,a_2,...,a_k\}$,  we create $k$ documents, each represented as  $\langle e, a_i \rangle: V_i$, where $V_i$ indicates all values of attribute $a_i$ in entity $e$.
 Following~\citep{Balog:2013:TCE}, we set the weights of MLM models to 0.2, and 0.8 for title and content fields (i.e., $a$ and $V$).
 We ran BM25 with parameters $k_1 = 1.2$ and $b = 0.8$ and Dirichlet smoothing with $\mu = 2000$ for LM and MLM-tc models.
In the MRF-based model, we set the parameters $\lambda_1 = 0.6$, $\lambda_2 = 0.2$, $\lambda_3 = 0.2$,  $\mu_1=0.5$, and $\mu_2=0.5$ (using parameter sweeps).
For LTR experiments, we used the CA implementation provided in the RankLib framework and set the number of random restarts to 3. We obtained the results using 5-fold cross validation, keeping attributes of the each query in the same fold.
We employed a two-tailed paired t-test ($\alpha = 0.05$) to measure statistical significance. Significant improvements over the best baseline model (i.e., MLM-tc) are marked with $^*$ in Table \ref{tab:results}.

\section{Results and Analysis}
Table \ref{tab:results} shows the comparison of the baseline and proposed methods. The NDCG@5, P@5, MRR, and MAP metrics are reported for all methods. The results show that the MRF-based model can significantly improve the baseline methods with respect to all metrics. This improvement can be explained by the fact that the baseline models rely only on exact matching of query and attribute terms, while the MRF-based model tries to score each attribute  by considering three similarities: entity-query, entity-attribute, and query-attribute.
The second observation is that the proposed LTR model (i.e., LTR/CA) improves the MRF-based model. This is expected, as 
the LTR method uses all the signals used by the MRF-based model (i.e., $f_1$, $f_2$, and $f_3$) as well as other features mentioned in Table~\ref{tab:features}. In addition, the LTR model uses an optimized combination of signals to rank attributes for a given query.

\begin{figure}[t]
	\includegraphics[width=.75\linewidth]{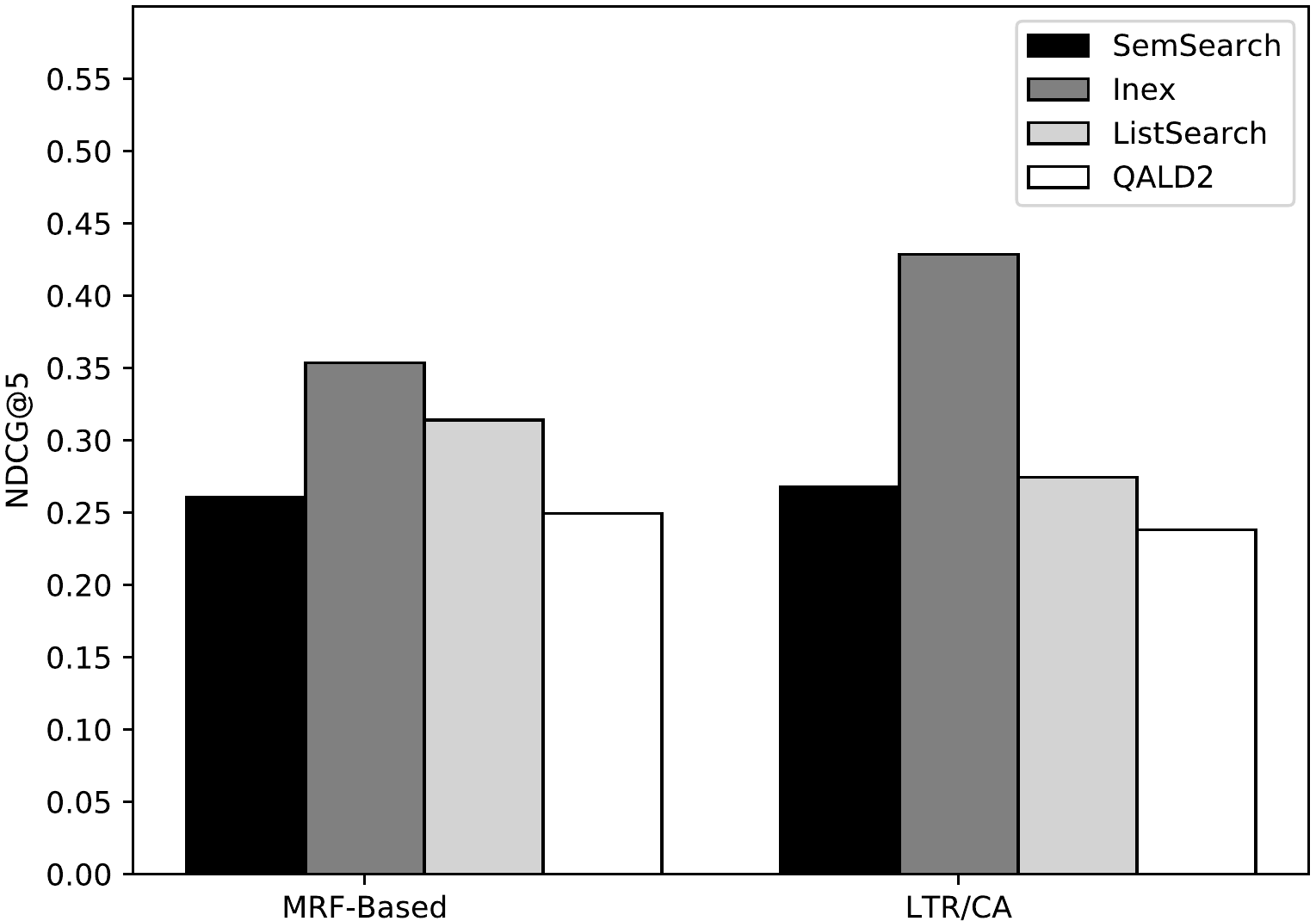}
	\vspace*{-0.8\baselineskip}
	\caption{Performance of MRF-based and LTR models for different query categories.}
	\label{fig:cat}
	\vspace*{-1.1\baselineskip}
\end{figure}

\begin{table}[t]
\shrink
  \centering
  \caption{Comparison of baselines and proposed models for attribute identification task.}
  \shrink
    \begin{tabular}{|l|cccc|}
    \toprule
    \textbf{Model} & NDCG@5 & P@5   & MRR   & MAP \\
    \midrule
	BM25  & 0.0467 & 0.0369 & 0.0749 & 0.0503 \\
	LM    & 0.0527 & 0.0371 & 0.0898 & 0.0618 \\
	MLM-tc   & 0.0803 & 0.0479 & 0.1168 & 0.0847 \\
    \midrule
    MRF-based  & 0.2844* & 0.1817*& 0.3618* & 0.2167* \\
    LTR/CA  & \textbf{0.3227*} & \textbf{0.2117*}& \textbf{0.3702*} & \textbf{0.3390*} \\
    \bottomrule
    \end{tabular}%
  \label{tab:results}%
\end{table}%
Figure \ref{fig:cat} indicates the comparison of proposed models for different query categories in our test collection. We observe that the retrieval performance (with respect to NDCG@5) on INEX-LD and ListSearch categories is higher than others. This can be explained by the fact that most of these queries are short and seeking for entities with a direct relation to the mentioned entity in the query. QALD queries, however, are complex and involve further understanding using natural language processing techniques.

We analyze the discriminative power of the features by comparing the ranking performance of each feature in isolation; i.e., using a single feature as a ranker. Table \ref{tab:fi} shows the results.
The third column indicates the NDCG@5 difference between single feature models and the model trained on all features. According to this table, features $f_5$ (Word2Vec similarity for linked-terms in q ) and $f_3$ (query-attribute similarity) have the most discriminative power. Both of these features consider the  similarity between query and entity attribute terms. 
Comparing features $f_4$ and $f_5$ with $f_6$ and $f_7$, we observe that query terms linked by entity linker systems usually contain more information about entity attributes of the queries than the not-linked terms.



\begin{table}[t]
	\centering
	\caption{Feature importance analysis.}
	\shrink
	\begin{tabular}{|c|cc|}
		\toprule
		Feature & NDCG@5 & $\Delta \%$ \\
		\midrule
		$f_1$--$f_7$  & 0.3227 & 0 \\
		\midrule
		$f_5$    & 0.2876 & -10.88\% \\
		$f_3$    & 0.2771 & -14.13\% \\
		$f_4$    & 0.2671 & -17.23\% \\
		$f_2$    & 0.1761 & -45.43\% \\
		$f_6$    & 0.0919 & -71.52\% \\
		$f_1$    & 0.0867 & -73.13\% \\
		$f_7$    & 0.0816 & -74.71\% \\
		\bottomrule
	\end{tabular}%
	\label{tab:fi}%
	\vspace*{-1.1\baselineskip}
\end{table}%

%% file: entity_attr_06.tex
\section{Conclusions and Future Work}
\label{sec:concl}

In this paper, we proposed the new task of entity attribute identification, which enables better understanding of search queries. 
We employed entity annotations of queries as a bridge to identify entity attributes of queries and proposed two methods to address this task.
 Since there is no available test collection for this task, we developed a new test collection based on an established test collection for entity retrieval. Using this collection, we examined our methods with a wide range of entity-bearing queries and showed that our models bring significant and substantial improvements over the baseline methods and are most effective for short relational queries. 
 For future, we plan to improve our model for complex natural language queries, and incorporate identified attributes of the query in the entity retrieval and question answering tasks.

